\documentstyle[12pt]{article}
\newcommand {\beq}{\begin{equation}}
\newcommand {\eeq}{\end{equation}}
\newcommand {\beqa}{\begin{eqnarray}}
\newcommand {\eeqa}{\end{eqnarray}}
\newcommand {\n}{\nonumber \\}

\begin{document}
\setlength{\oddsidemargin}{0cm}
\setlength{\baselineskip}{7mm}
\begin{titlepage}
 \renewcommand{\thefootnote}{\fnsymbol{footnote}}
$\mbox{ }$
\begin{flushright}
\begin{tabular}{l}
hep-th/0203277 \\
KEK-TH-804 \\
SAGA-HE-186 \\
March 2002
\end{tabular}
\end{flushright}

~~\\
~~\\
~~\\

\vspace*{0cm}
    \begin{Large}
       \vspace{2cm}
       \begin{center}
         {Orbifold Matrix Model}      \\
       \end{center}
    \end{Large}

  \vspace{1cm}

\begin{center}
           Hajime A{\sc oki}$^{1)}$\footnote
           {
e-mail address : haoki@cc.saga-u.ac.jp},
           Satoshi I{\sc so}$^{2)}$\footnote
           {
e-mail address : satoshi.iso@kek.jp}, and
           Takao S{\sc uyama}$^{2)}$\footnote
           {
e-mail address : tsuyama@post.kek.jp} \\
        $^{1)}$ {\it Department of Physics, Saga University,
Saga 840-8502, Japan}\\
        $^{2)}$ {\it High Energy Accelerator Research Organisation
             (KEK), \\
             {\it Tsukuba, Ibaraki 305-0801, Japan} }\\
\end{center}

\vfill

\begin{abstract}
\noindent
We study a matrix model describing type IIB superstring
in orbifold backgrounds.
We particularly consider a ${\bf C}^3/{\bf Z}_3$ orbifold model
whose six dimensional transverse space is orbifolded by
${\bf Z}_3$ discrete symmetry.
This model is chiral and has $d=4$ ${\cal N}=1$
supersymmetry of Yang-Mills type as well as an inhomogeneous
supersymmetry specific to matrix models.
We calculate one-loop effective action around some backgrounds,
and the result
can be interpreted as interactions mediated by massless particles
in IIB supergravity in orbifold background, if the background is
in the Higgs branch.
If the background is in the Coulomb branch, the dynamics
is governed by the reduced model of $d=4$ super Yang-Mills theory,
which might be interpreted as exchange of massless particles in the twisted
sector.
But the perturbative calculation does not reproduce the
supergravity result.
We also show that this model
with a large Higgs vacuum expectation value
becomes IIB (IKKT) matrix model.

\end{abstract}

\vfill
\end{titlepage}
\vfil\eject

\section{Introduction}
\setcounter{equation}{0}
\setcounter{footnote}{0}
A large $N$ reduced model has been proposed as a nonperturbative
formulation of type IIB superstring theory\cite{IKKT}.
It is defined by the action\footnote{In this paper, we take a
metric convention $\eta_{IJ}=\mbox{diag}(-1,1,\cdots,1)$.}:
\beq
S  =  -{1\over g^2}Tr({1\over 4}[A_{I},A_{J}][A^{I},A^{J}]
+{1\over 2}\bar{\psi}\Gamma ^{I}[A_{I},\psi ]) .
\label{action}
\eeq
It is a large $N$ reduced model \cite{RM}
of ten dimensional super Yang-Mills
theory.
Here $\psi$ is a ten dimensional Majorana-Weyl spinor field, and
$A_{I}$ and $\psi$ are $N \times N$ Hermitian matrices.
It is formulated in a manifestly covariant way which enables us
to study the nonperturbative issues of superstring theory.
In fact we can, in principle, predict the dimensionality of spacetime,
the gauge group and the matter contents by solving this
model\cite{review}.
\par
Among  several issues on the matrix formulation of superstring,
one  is to answer which model should be regarded as the most
fundamental one. This question can be compared to the
universality in the lattice gauge theory.
In this case, the most important symmetry we have to keep
is the gauge symmetry and we know that various models
with the gauge invariance are in the same universality class.
In the case of matrix models, maximal supersymmetry
may play the same role but it is not yet certain.
Hence, it is meaningful to consider various matrix models,
investigate their dynamics and make connections between them.
If matrix models can describe the dynamics of space-time,
we should be able to generate various space-time
backgrounds from the same model, and in this respect universality
in matrix models will be related to the background
independence.
\par
Another related issue is to construct a phenomenologically
interesting model
with chiral fermions in $d=4$.
In string theory, there are several mechanisms
to generate four-dimensional chiral fermions,
including compactification with non-zero index of Dirac
operator, D-branes wrapped on intersecting cycles, etc \cite{Uranga}.
In matrix models, we may similarly be able to
consider corresponding mechanisms
but no investigations have been performed yet.
\par
In this paper we study a matrix model
in ${\bf C}^3/{\bf Z}_3$ orbifold  background \cite{Kachru}
\cite{MatrixOrbifold}.
This model can be obtained from  IIB matrix model by
imposing some constrains.
In particular we consider a model with four-dimensional
${\cal N}=1$ Yang-Mills type supersymmetry.
In matrix models, there is also an inhomogeneous
supersymmetry as well,
and the model  has totally $d=4$ ${\cal N}=2$
supersymmetries.
This model is chiral since the chiral multiplets
are in bi-fundamental representation of the
gauge group.
There are two types of classical solutions,
corresponding to the Higgs branch and the Coulomb
branch, respectively.
The Higgs branch solution forms a triplet
under ${\bf Z}_3$ transformation and there are three
images in ${\bf C}^3$ plane whose positions in
$d=4$ directions are the same.
On the other hand, the Coulomb branch solution
is restricted on the orbifold fixed point and
cannot move into ${\bf C}^3$ direction. The Coulomb branch
solution is continuously connected to the Higgs branch solution
at the orbifold singularity. When the Higgs branch
solution comes on the orbifold singularity, the
three images are liberated from each other and
they can move independently into $d=4$ directions.
As BPS solutions, the former represents  ordinary D-branes
with three mirror images by ${\bf Z}_3$,
and when they
lie on the singularity they can move freely into the
four dimensional direction as fractional branes.
\par
We calculate one-loop effective action of the
orbifold matrix model to clarify its correspondence
to the interaction in type IIB  supergravity in the
orbifold background \cite{DO}. Firstly we show that
 we can reproduce
the correct supergravity interaction for
classical solutions in the Higgs branch in orbifold background.
Secondly the dynamics of
fractional branes in the Coulomb branch
is shown to be governed by the
reduced model of $d=4$ super Yang-Mills
theory instead of $d=10$ Yang-Mills theory.
In string theory,
fractional branes \cite{Fbrane} have a charge of
RR fields in the twisted sector and the interaction
between them should have four-dimensional
behaviour in the long distance.
In this respect, it is qualitatively consistent with the string
picture. But perturbatively the reduced model of $d=4$
super Yang-Mills theory does not
reproduce the correct interactions
mediated by massless fields in the twisted
sector. (The induced interaction is inversely
proportional to
the fourth power of the distance instead of
the expected  quadratic power.)
This discrepancy will be a limitation of
perturbative calculations and it
 is not certain whether incorporation
of nonperturbative effects can help it.

This model has smaller supersymmetries compared with
IIB matrix model and it is not apparent
if it is in the same universality class as
 IIB matrix model. But if it is, the orbifold
matrix model must be
obtained from IIB matrix model through
condensation of some fields and so must be the reverse.
These relations will also shed light on the issue
how to realize four-dimensional chiral fermions from IIB matrix model.
In this paper, we show that the orbifold matrix model
in the Higgs phase with a large expectation
value becomes  IIB matrix model.
\par
The organisation of the paper is as follows.
In section 2, we give a brief review of IIB
matrix model.
In section 3, we construct  orbifold
matrix models, and particularly consider
a model with ${\cal N}=1$ supersymmetry of
Yang-Mills type in $d=4$.
In section 4, we analyse this orbifold matrix
model. First we list several classical solutions
with or without noncommutativity of backgrounds.
Then we calculate one-loop effective action
around these solutions and show that
the result is partly consistent with the string theory.
In section 5, we show that this model
with a large Higgs vacuum expectation value
becomes IIB matrix model.
Section 6 is devoted to conclusions and discussions.

\setcounter{equation}{0}
\section{Brief Review of IIB Matrix Model}
Action of the  IIB matrix model is defined in (\ref{action}).
This model has ${\cal N}=2$ ten-dimensional supersymmetry if we regard the
distribution of eigenvalues as space-time.
Since this matrix model action can be obtained by reducing
the original ten dimensional space-time to a single point,
the action has the remnant of the supersymmetry of super Yang-Mills
type:
\beqa
\delta^{(1)}\psi &=& \frac{i}{2}
                     [A_{I},A_{J}]\Gamma^{IJ}\epsilon ,\n
\delta^{(1)} A_{I} &=& i\bar{\epsilon }\Gamma^{I}\psi .
\label{Ssym1}
\eeqa
In addition to this homogeneous symmetry, there is an inhomogeneous
fermionic symmetry
\beqa
\delta^{(2)}\psi &=& \xi ,\n
\delta^{(2)} A_{I} &=& 0.
\label{Ssym2}
\eeqa
If we take a linear combination of $\delta^{(1)}$ and $\delta^{(2)}$ as
\beqa
\tilde{\delta}^{(1)}&=&\delta^{(1)}+\delta^{(2)}, \n
\tilde{\delta}^{(2)}&=&i(\delta^{(1)}-\delta^{(2)}),
\label{susycombine}
\eeqa
we obtain the ${\cal N}=2$ supersymmetry algebra:
\beqa
(\tilde{\delta}^{(i)}_{\epsilon}\tilde{\delta}^{(j)}_{\xi}
    -\tilde{\delta}^{(j)}_{\xi}\tilde{\delta}^{(i)}_{\epsilon})\psi   &=&0
,\n
(\tilde{\delta}^{(i)}_{\epsilon}\tilde{\delta}^{(j)}_{\xi}
    -\tilde{\delta}^{(j)}_{\xi}\tilde{\delta}^{(i)}_{\epsilon})A_{I}&=&
                                 2i\bar{\epsilon}\Gamma^{I}\xi
\delta_{ij}.
\eeqa
The ${\cal N}=2$ supersymmetry is a crucial element of superstring theory.
It imposes strong constraints on the spectra of particles.
Furthermore it determines the structure of the interactions uniquely in
the light-cone string field theory\cite{FKKT}.
The IIB matrix model is a nonperturbative formulation which possesses
such a symmetry. Therefore it has a very good chance to capture the
universality class of IIB superstring theory.
These symmetry considerations force us to interpret
the eigenvalues of $A_{I}$ as the space-time coordinates.

The simplest classical solutions are
those where all bosonic matrices are commutable and
 simultaneously diagonalisable.
Since the distributions of the eigenvalues
determine the extent and the dimensionality
of spacetime, the structure of spacetime can be dynamically determined by
the theory;
Indeed, spacetime exits as a single bunch
and no single eigenvalue can escape from the rest\cite{AIKKT}.
Numerical simulations and analysis with various approximations
have indicated that
the Lorentz symmetry might be spontaneously broken \cite{AIKKT,Nishimura}.

The next simplest solutions are
those with  non-zero but c-number commutator
$[A_{I}, A_{J}]=i \theta_{IJ}{\bf 1}_{N \times N}$.
These solutions
correspond to BPS-saturated backgrounds. Indeed,
the solutions are invariant under transformations if we set
$\xi$ equal to $\pm \frac{1}{2} \theta_{IJ} \Gamma^{IJ}\epsilon$
in the $\cal{N}$=2 supersymmetry (\ref{Ssym1}) and (\ref{Ssym2}).
By expanding the matrix around this classical solution,
we can obtain a supersymmetric noncommutative gauge theory \cite{NCMM}.

Finally in this section, we comment on
the calculation of the one-loop effective action
in IIB matrix model and its interpretation in type IIB supergravity.
In \cite{IKKT}, it was shown that when the bosonic matrices are
expanded around backgrounds having a block-diagonal form
\begin{eqnarray}
A_{\mu} = X_{\mu} =\left(
\begin{array}{cc}
d_{\mu(1)}+X_{\mu(1)} & \\
 &  d_{\mu(2)}+X_{\mu(2)}
\end{array}
\right),
\end{eqnarray}
the one-loop effective action becomes
\begin{eqnarray}
W^{(12)}&=&\frac{1}{4(d^{(1)}-d^{(2)})^8}\n
         & & ([
          (-4n_2 Tr(\tilde{f}^{(1)}_{\mu\nu}\tilde{f}^{(1)}_{\nu\lambda}
                  \tilde{f}^{(1)}_{\lambda\rho}\tilde{f}^{(1)}_{\rho\mu})
             -8n_2 Tr(\tilde{f}^{(1)}_{\mu\nu}\tilde{f}^{(1)}_{\lambda\rho}
                  \tilde{f}^{(1)}_{\mu\rho}\tilde{f}^{(1)}_{\lambda\nu})\n
         & &+2n_2 Tr(\tilde{f}^{(1)}_{\mu\nu}\tilde{f}^{(1)}_{\mu\nu}
\tilde{f}^{(1)}_{\lambda\rho}\tilde{f}^{(1)}_{\lambda\rho})
            +n_2 Tr(\tilde{f}^{(1)}_{\mu\nu}\tilde{f}^{(1)}_{\lambda\rho}
                  \tilde{f}^{(1)}_{\mu\nu}\tilde{f}^{(1)}_{\lambda\rho})\n
   && -48 Tr(\tilde{f}_{\mu \nu}^{(1)})
   Tr(\tilde{f}^{(2)}_{\mu \lambda} \tilde{f}^{(2)}_{\lambda \rho}
    \tilde{f}^{(2)}_{\rho \mu})
  + 12 Tr(\tilde{f}_{\mu \nu}^{(1)})
    Tr(\tilde{f}^{(2)}_{\mu \nu}
    \tilde{f}^{(2)}_{\lambda \rho}
    \tilde{f}^{(2)}_{\lambda\rho})
             + (1 \leftrightarrow 2) ] \n
             & &-48Tr(\tilde{f}^{(1)}_{\mu\nu}\tilde{f}^{(1)}_{\nu\lambda})
              Tr(\tilde{f}^{(2)}_{\mu\rho}\tilde{f}^{(2)}_{\rho\lambda})
            +6Tr(\tilde{f}^{(1)}_{\mu\nu}\tilde{f}^{(1)}_{\mu\nu})
Tr(\tilde{f}^{(2)}_{\lambda\rho}\tilde{f}^{(2)}_{\lambda\rho}))\n
         & &  +O((1/(d^{(1)}-d^{(2)})^9)
\label{bbIIB}
\end{eqnarray}
where $f_{\mu \nu}^{(i)}=i[X_{\mu}^{(i)},X_{\nu}^{(i)}]$
and we keep terms which vanish for finite $N$.
(These terms play an important role in calculating interactions
between objects with charges of antisymmetric fields such as D-branes.)
Observing the tensor structures, we find the exchanges of
massless particles in IIB supergravity corresponding to
graviton, scalars and antisymmetric fields.
In this paper, we perform similar calculations in the matrix
model in orbifold backgrounds.

\setcounter{equation}{0}
\section{${\bf C}^3/{\bf Z}_3$ Orbifold Matrix Model}
\subsection{${\bf Z}_3$ Orbifolding}
We now construct  matrix models in ${\bf C}^3/{\bf Z}_3$ orbifold
background \cite{MatrixOrbifold}.
The IIB matrix model
consists of ten hermitian bosonic matrices $A_{I}$ ($I=0,\cdots,9$)
and
$d=10$ Majorana-Weyl fermionic matrices with sixteen components.
We take complex coordinates for the six-dimensional transverse space
and write them as
$B_i$ (i=1,2,3),
\begin{equation}
B_1=A_4 + iA_5, \ \ \ B_2=A_6 + iA_7, \ \ \ B_3=A_8 + iA_9.
\end{equation}
We characterise  the action of ${\bf Z}_3$ group  by three
integers $(a_1,a_2,a_3)$.
The ${\bf Z}_3$ transformation is
defined to act on these complex coordinates as
$B_i \rightarrow \omega^{a_i} B_i$.
Different choices of these integers lead to different models
with a different number of unbroken supersymmetries in $d=4$.
\par  From the D-instanton point of view, since
there are three images of each D-instanton,
we have to extend the size of the matrices three times
to represent these three mirror images.
The ${\bf Z}_3$ group rotates
these three images.
We write hermitian matrices
as tensor products of $3 \times 3$
 matrices on which ${\bf Z}_3$ acts and the rest.
The  $3 \times 3$ matrices can be expanded
in  terms of 't Hooft matrices,
\begin{equation}
U=\left( \begin{array}{ccc}
  1 & & \\
    & \omega & \\
    & & \omega^2
\end{array}
\right), \ \ \
V= \left( \begin{array}{ccc}
   0 & 0 &  1 \\
   1 &  0 & 0 \\
   0 & 1 & 0
\end{array}
\right),
\label{thooftUV}
\end{equation}
where $UV = \omega VU$.
In a representation where
the coordinates of each mirror image of D-instantons
is diagonalized, the ${\bf Z}_3$ transformation
interchanges these diagonal blocks,
and is expressed
as $M \rightarrow VMV^{\dagger}$,
where $M$ is $A_{I}$ or $\psi$.
While this representation is geometrically clear,
for later convenience,
we take another representation
where the ${\bf Z}_3$ transformation is given as
$M \rightarrow UMU^{\dagger}$.
\par
We now impose ${\bf Z}_3$ invariance conditions on matrices.
The ${\bf Z}_3$ condition for the gauge field ($A_{\mu}$, $\mu=0 ,,,3$)
is
\begin{equation}
A_{\mu} = U A_{\mu} U^{\dagger}. \label{Z3-1}
\end{equation}
The off-diagonal blocks are projected out
and only the diagonal blocks $A_{\mu (i,i)}$, $i=1,2,3$, of
$3 \times 3$ matrices survive after this ${\bf Z}_3$ projection.
If the size of each block matrix is $N \times N$,
these gauge fields are associated with $U(N)^3$ symmetry.
The ${\bf Z}_3$ invariance condition for the complex
fields $B_i$ is given by
\begin{equation}
B_i = \omega^{a_i} \ U B_i U^{\dagger}  \label{Z3-2}
\end{equation}
and  is written as
$a_i + i-j =0 (\mbox{ mod } 3)$ for $(i,j)$ blocks.
\par
Fermionic matrices are similarly projected out by
this ${\bf Z}_3$ transformation.
A convenient way to pick up ${\bf Z}_3$ invariant components
out of the original $d=10$ Majorana-Weyl fermionic matrix
$\psi$ is to write it in lower dimensional Weyl-basis.
Since $d=10$ chirality operator $\Gamma=\Gamma^{01 \cdots 9}$
is written as a product of lower dimensional chirality operators
as
$\Gamma =(i\Gamma^{0123})(i\Gamma^{45})(i\Gamma^{67})(i\Gamma^{89})$,
a positive chirality fermion $\psi$ in $d=10$ can be
decomposed as a sum of the fermions
$
\psi=  \sum_{i=0}^3 (\psi_{L,i} +\psi_{R,i})
$
in the following table.
\begin{equation}
\begin{array}{|c||c|c|c|c||c|c|c|} \hline
\Gamma &i\Gamma^{0123} &i\Gamma^{45}&i\Gamma^{67} & i\Gamma^{89}&
& \mbox{Majorana} & b_i
\\ \hline \hline
+ &+&+&+&+&\psi_{L,0}&\psi_0 & (a_1+a_2+a_3)/2\\ \cline{2-8}
 &+&+&-&-&\psi_{L,1}&\psi_1 & (a_1-a_2-a_3)/2 \\ \cline{2-8}
 &+&-&+&-& \psi_{L,2}&\psi_2 & (-a_1+a_2-a_3)/2 \\ \cline{2-8}
 &+&-&-&+& \psi_{L,3}&\psi_3 & (-a_1-a_2+a_3)/2 \\ \cline{2-8}
 &-&-&-&-& \psi_{R,0}&(\psi_{0})^c & (-a_1-a_2-a_3)/2 \\ \cline{2-8}
 &-&-&+&+& \psi_{R,1}&(\psi_{1})^c  & (-a_1+a_2+a_3)/2 \\ \cline{2-8}
 &-&+&-&+&  \psi_{R,2}&(\psi_{2})^c & (a_1-a_2+a_3)/2 \\ \cline{2-8}
 &-&+&+&-&  \psi_{R,3}&(\psi_{3})^c & (a_1+a_2-a_3)/2 \\ \hline
\end{array}
\label{fermiontable}
\end{equation}
Since the $d=10$ charge conjugation interchanges left and right as
$(\psi_{L,i})^c=(\psi^c)_{R,i}$,
a $d=10$ Majorana-Weyl fermion $\psi$ contains
four independent complex degrees of freedom $\psi_{L,i}$,
and the right-handed fermions  are related to the left as
$\psi_{R,i}=(\psi_{L,i})^c$.
In the following we write $\psi_{L,i}$ without index $L$,
and $\psi_{R,i}$ as its charge conjugation. Hence
the fermionic $3N \times 3N$ matrix $\psi$ is decomposed as
\begin{equation}
\psi=\sum_{i=0}^3 (\psi_i^A +(\psi_i^A)^c) T^A
\end{equation}
where $T^{A}$'s  are generators of $3N \times 3N$ hermitian matrices.
Since the ${\bf Z}_3$ transformation acts on each complex plane
${\bf C}$ in ${\bf C}^3$, ${\bf Z}_3$ invariant
condition depends on how each fermion transforms under rotations
in each ${\bf C}$ plane. Hence,
according to the eigenvalues of
$i\Gamma^{45}$, $i\Gamma^{67}$ and $i\Gamma^{89}$,
each of the fermionic matrices is subject to  the following
${\bf Z}_3$
invariance condition:
\begin{equation}
\psi_i = \omega^{b_i} \ U \psi_i U^{\dagger},
\label{Z3-3}
\end{equation}
where the charges $b_i$'s are calculated from $a_i$'s
as (\ref{fermiontable}).
Invariant blocks of $3 \times 3$ matrices
satisfy $b_i +i-j=0 (\mbox{ mod } 3)$.

\par
An action of the orbifold matrix model is given from the
 IIB matrix model action (\ref{action})
by restricting the matrices by the $Z_{3}$
constraints (\ref{Z3-1}), (\ref{Z3-2}) and (\ref{Z3-3}).
First we rewrite the IIB matrix model action (\ref{action})
in terms of the $d=4$ gauge field and the complex coordinate fields.
The bosonic part of IIB matrix model action becomes
\begin{equation}
S_b = -\frac{1}{4 g^2} Tr
 \left(  [A_{\mu}, A_{\nu}]^2
+ 2 \sum_{i=1}^3 [A_{\mu}, B_i][A^{\mu}, B^{\dagger}_i]
 + \frac{1}{2} \sum_{i,j=1}^3 ([ B_i, B^{\dagger}_j][B^{\dagger}_i, B_j]
+[B_i,B_j][B^{\dagger}_i,B^{\dagger}_j])
\right).
\label{bosonaction1}
\end{equation}
The $Tr$ is a trace over $3N \times 3N$ matrices.
The bosonic part of the orbifold matrix model action
is given by (\ref{bosonaction1}) where matrices are constrained
by the conditions (\ref{Z3-1}), (\ref{Z3-2}).
The potential term for $B_i$ field can be
written as
\begin{equation}
- \frac{1}{2} Tr(  |\sum_i [B_i,B_i^{\dagger}]|^2 +
 2 \sum_{i,j} |[B_i, B_j]|^2 )
\end{equation}
and each term in this expression comes from
the integration of D term and  F term respectively
in the superspace formalism.
Comparing to the  $N \times N$  IIB matrix model (\ref{action}),
the  $3N \times 3N$ orbifold matrix model
has three times as many as matrices.
\par
The fermionic part of the IIB matrix model action (\ref{action})
is rewritten in terms of the four
dimensional fermions in the table (\ref{fermiontable}) as
\begin{equation}
S_f= - \frac{1}{2 g^2} Tr \left(
\sum_{i=0}^{3}
 \overline{\psi_i} \Gamma^{\mu} [A_{\mu},\psi_i]
+ 2 \sum_{i=1}^{3} \overline{(\psi_i)^c} \bar{\Gamma}^{(i)}
[B^{\dagger}_i,\psi_0]
+ \sum_{i,j,k=1}^3 |\epsilon_{ijk}|  \overline{(\psi_i)^c} \Gamma^{(j)}
[B_j, \psi_k] + h.c.
\right)
\label{fermionaction1}
\end{equation}
where we have defined complex gamma matrices
\begin{equation}
\Gamma^{(1)} \equiv \frac{1}{2} (\Gamma^4-i\Gamma^5), \ \
\bar{\Gamma}^{(1)} \equiv \frac{1}{2} (\Gamma^4 + i\Gamma^5),
\end{equation}
and similarly for $\Gamma^{(2)}$ and $\Gamma^{(3)}$.
The fermionic part of the orbifold matrix model is given by
(\ref{fermionaction1}) where fermionic matrices are restricted by
the condition (\ref{Z3-3}).
\par
By ${\bf Z}_3$ orbifolding, the original symmetry of IIB matrix model
with size $3N \times 3N$ is generally
reduced from $SO(9,1) \times U(3N)$ to
$SO(3,1) \times U(N)^3$.
${\bf Z}_3$ invariant fermion
fields transform as bi-fundamental representations
under the unbroken $U(N)^3$ gauge symmetry
and the resulting gauge theory
becomes a so-called quiver gauge theory.
A subgroup of the transverse $SO(6)$ symmetry
is also unbroken. It depends on the charge assignment $a_i.$
A model in the next
subsection has unbroken $U(3)$ symmetry,
which rotate three chiral superfields.
\par
As for supersymmetry, we  decompose the supersymmetry
parameter $\epsilon$ and $\xi$ of (\ref{Ssym1}) and (\ref{Ssym2})
into the lower dimensional Weyl basis in the table (\ref{fermiontable}).
If some of these supersymmetry parameters are invariant under the
${\bf Z}_3$
transformation, they transform ${\bf Z}_3$ invariant fields among
themselves.
Therefore, the number of zero charges of $b_i (i=0,\cdots,3)$
is the number of unbroken four dimensional supersymmetries.
While in matrix models, we have inhomogeneous
supersymmetries (\ref{Ssym2})
as well as the Yang-Mills type (\ref{Ssym1}),
in this paper, we will name the model by
the number of Yang-Mills type supersymmetry.

\subsection{${\cal N}=1$ Model}
We particularly consider a model
with  ${\bf Z}_3$ charge assignment
$a_i=2$ for $i=1,2,3$, and accordingly  $b_0=0, b_i=2 (i=1,2,3)$.
The ${\bf Z}_3$ invariance conditions
(\ref{Z3-1})(\ref{Z3-2})(\ref{Z3-3}) can be
satisfied if we restrict the matrices to the following form
\begin{eqnarray}
&& A_{\mu} =\left( \begin{array}{ccc}
  A_{\mu}^{(1)}  &0 &0 \\
   0 &  A_{\mu}^{(2)} &0 \\
   0 &0 &  A_{\mu}^{(3)}
\end{array}
\right), \ \ \
B_{i} = \left( \begin{array}{ccc}
   0 & 0 &  B_{i}^{(2)} \\
   B_{i}^{(3)} &  0 & 0 \\
   0 & B_{i}^{(1)} & 0
\end{array}
\right),
\ \ \
B^{\dagger}_{i} = \left( \begin{array}{ccc}
   0 & B_{i}^{(3) \dagger} &  0 \\
   0 &  0 & B_{i}^{(1)\dagger} \\
   B_{i}^{(2)\dagger} & 0 & 0
\end{array}
\right), \nonumber \\
&& \psi_0= \left( \begin{array}{ccc}
  \psi_0^{(1)}  & 0&0 \\
    0&  \psi_0^{(2)} &0 \\
    0& 0&  \psi_0^{(3)}
\end{array}
\right),
\ \ \
\psi_{i} = \left( \begin{array}{ccc}
   0 & 0 &  \psi_{i}^{(2)} \\
   \psi_{i}^{(3)} &  0 & 0 \\
   0 & \psi_{i}^{(1)} & 0
\end{array}
\right), \nonumber \\
&& (\psi_0)^c= \left( \begin{array}{ccc}
  (\psi_0)^{c(1)}  &0 &0 \\
    0&  (\psi_0)^{c(2)} &0 \\
    0&0 &  (\psi_0)^{c(3)}
\end{array}
\right),
\ \ \
(\psi_i)^c = \left( \begin{array}{ccc}
   0 & (\psi_i)^{c(3)} &  0 \\
   0 &  0 & (\psi_i)^{c(1)} \\
   (\psi_i)^{c(2)} & 0 & 0
\end{array}
\right).
\label{blockrep}
\end{eqnarray}
Inserting these expansions into (\ref{bosonaction1}) and
(\ref{fermionaction1}),
we can obtain the explicit form of the action for the orbifold matrix model.

As we will see at the end of this subsection,
this model has ${\cal N}=1$ Yang-Mills type supersymmetry.
$A_{\mu}$ and $\psi_0$ fields form a vector multiplet while
$B_i$ and $\psi_i$ fields make three chiral multiplets.
The left and right handed fermions in the chiral multiplets
($\psi_i$ and $(\psi_i)^c$) are in  different
bifundamental representations of $U(N)^3$ and hence the model
is chiral in $d=4$
and phenomenologically attractive.
It is also interesting to consider other models with
different charge assignments, symmetries, and chirality.
\par
We can also write
the ${\bf Z}_3$ invariant fields as
\begin{eqnarray}
&& A_{\mu} = \sum_{a=0,1,2} A_{\mu}^a \otimes U^a, \ \ \
B_i = \sum_{a=0,1,2} B_i^a \otimes (U^a V), \\
&& \psi_0 = \sum_{a=0,1,2} \psi_{0}^a \otimes U^a, \ \ \
\psi_i = \sum_{a=0,1,2} \psi_{i}^a \otimes (U^a V),
\label{UVexpansion}
\end{eqnarray}
where $U$ and $V$ are the 't Hooft matrices (\ref{thooftUV}).
Since $A_\mu$'s are hermitian,
$(A_{\mu}^a)^{\dagger}=A_{\mu}^{-a}$.
Charge conjugated and
hermitian conjugated fields are expanded as
\begin{eqnarray}
&& B_i^{\dagger} = \sum_{a=0,1,2} B_i^{a \dagger}
\otimes (V^{\dagger }U^{-a}), \nonumber \\
&& (\psi_0)^c = \sum_{a=0,1,2} (\psi_{0}^a)^c \otimes U^{-a}, \ \ \
(\psi_i)^c = \sum_{a=0,1,2} (\psi_{i}^a)^c \otimes (V^{\dagger }U^{-a}),
\nonumber \\
&& \overline{\psi_i} =\sum \overline{\psi_i^a}
\otimes (V^{\dagger }U^{-a}),
\ \ \ \overline{(\psi_i)^c} =  \sum_{a=0,1,2} \overline{(\psi_{i}^a)^c}
\otimes (U^a V).
\end{eqnarray}
These expansions are more useful than the block representation
(\ref{blockrep})
when we investigate this model in the next section.
\par
Since  only $b_0$ vanishes,
there is  one $d=4$ Yang-Mills type supersymmetry in this model.
Inhomogeneous  fermionic transformations  (\ref{Ssym2})
also make the orbifold matrix model action invariant only
if the supersymmetry parameter $\xi$ is the $\psi_0$ type.
The Yang-Mills type supersymmetry of this
orbifold matrix model is given by
\begin{eqnarray}
\delta^{(1)} \psi_0 &=& \frac{i}{2}([A_{\mu},A_{\nu}] \Gamma^{\mu \nu}
+[B_i, B_i^{\dagger}]) \epsilon_0 ,
 \\
\delta^{(1)} \psi_i  &=& \frac{i}{2} (|\epsilon_{ijk}|[B_j^{\dagger},
B_k^{\dagger}] \bar{\Gamma}^{(j)} \bar{\Gamma}^{(k)} \epsilon_0
+ 2 [A_{\mu}, B_i] \Gamma^{\mu} \Gamma^{(i)} (\epsilon_0)^c
),
 \\
\delta^{(1)} A_{\mu} &=& i \overline{\epsilon_0} \Gamma^{\mu}\psi_0
+ i \overline{(\epsilon_0)^c} \Gamma^{\mu} (\psi_0)^c ,
\\
\delta^{(1)} B_i &=& 2i
\overline{(\epsilon_0)^c}  \bar{\Gamma}^{(i)}  \psi_i .
\end{eqnarray}
The other supersymmetry is
\begin{equation}
\delta^{(2)} \psi_0 = \xi_0.
\end{equation}
Here $\epsilon_0$ and $\xi_0$ are supersymmetry parameters restricted to
the $\psi_0$ type component in the table (\ref{fermiontable}).
Combining them similarly to (\ref{susycombine}), they
become $d=4$ ${\cal N}=2$ space-time supersymmetry whose commutator
gives translation of $A_{\mu}$ ($\mu=0,,3$).
In matrix models,  bosonic fields are interpreted as space-time
coordinates and hence the above supersymmetry can be identified
with the ${\cal N}=2$ four-dimensional space-time supersymmetry.
In this paper, however, we call this model ${\cal N}=1$
orbifold matrix model,
counting only the number of Yang-Mills type supersymmetries.

\setcounter{equation}{0}
\section{Investigations of ${\cal N}=1$ Orbifold Matrix Model}
\subsection{Classical Solutions}
We now investigate
 the orbifold matrix model, especially ${\cal N}=1$ type model.
Equations of motion  can be obtained from
the equations of motion of IIB matrix model
$ [A_I,[A_I,A_J]]=0 $ by restricting $A_I$
to the ${\bf Z}_3$ invariant parts.
This is
because the projected-out fields by ${\bf Z}_3$ symmetry
always enter in the action with another projected-out field.
\par
We first search for commutable solutions $[A_{I},A_J]=0.$
This equation can be written
in terms of the ${\bf Z}_3$ invariant
components in the block representation (\ref{blockrep}) as
\begin{eqnarray}
 &&  [A_{\mu}^{(\alpha)},A_{\nu}^{(\alpha)}] =0,
\label{eom-1}
 \\
 &&  A_{\nu}^{(1)}B_{i}^{(2)} - B_{i}^{(2)}A_{\nu}^{(3)} =
A_{\nu}^{(2)}B_{i}^{(3)} - B_{i}^{(3)}A_{\nu}^{(1)}=
A_{\nu}^{(3)}B_{i}^{(1)} - B_{i}^{(1)}A_{\nu}^{(2)}=0,
\label{eom-2}
 \\
 &&  B_{i}^{(1)} B_{j}^{(3)} - B_{j}^{(1)} B_{i}^{(3)}=
 B_{i}^{(2)} B_{j}^{(1)} - B_{j}^{(2)} B_{i}^{(1)}=
 B_{i}^{(3)} B_{j}^{(2)} - B_{j}^{(3)} B_{i}^{(2)}=0,
\label{eom-3} \\
 &&   B_{i}^{(1)} B_{j}^{(1)\dagger}- B_{j}^{(2)\dagger} B_{i}^{(2)}=
  B_{i}^{(2)} B_{j}^{(2)\dagger}- B_{j}^{(3)\dagger} B_{i}^{(3)}=
  B_{i}^{(3)} B_{j}^{(3)\dagger}- B_{j}^{(1)\dagger} B_{i}^{(1)}=0.
\label{eom-4} \end{eqnarray}
\par
There are two types of solutions to these equations.
The first type
has generally non-vanishing v.e.v. of $B_i$ and
corresponds to the
Higgs branch of super Yang-Mills theory. In this case,
$A_{\mu}^{(\alpha)}$ must be independent of the block index ($\alpha$)
and has the form
\begin{equation}
 A_{\mu}^{(cl)} =\left( \begin{array}{ccc}
  X_{\mu}  & & \\
    & X_{\mu} & \\
    & & X_{\mu}
\end{array}
\right),
\label{sol1A}
\end{equation}
where $X_{\mu}$ are $N \times N$ diagonal matrices.
The equation (\ref{eom-2}) can be satisfied if
$B_{i}^{(\alpha)}$ is a diagonal matrix.
To satisfy the last two equations
(\ref{eom-3}) and (\ref{eom-4}), we further assume that $B_{i}^{(\alpha)}$
is independent of the block index ($\alpha$) and has the form
\begin{equation}
 B_{i}^{(cl)} = \left( \begin{array}{ccc}
   0 & 0 &  B_{i}^{(2)(cl)} \\
   B_{i}^{(3)(cl)} &  0 & 0 \\
   0 & B_{i}^{(1)(cl)} & 0
\end{array}
\right) = \left( \begin{array}{ccc}
   0 & 0 &  Z_{i} \\
   Z_{i} &  0 & 0 \\
   0 & Z_{i} & 0
\end{array}
\right)
\label{sol1B}
\end{equation}
where $Z_i$ is an $N \times N$ diagonal matrix.
More generally,
phases $\theta^{\alpha}$ of
$B_i^{(\alpha)(cl)} $ can be put differently  as
$B_i^{(\alpha)(cl)} = e^{i \theta^{\alpha}} Z_i$
(where $\theta^{\alpha}$ is an $N \times N$ diagonal matrix).
However, since they
transform under $U(1)^{3N}$ subgroup
of the $U(N)^3$ symmetry as
\begin{equation}
 \theta_{\alpha} \rightarrow \theta_{\alpha} +
 (\phi_{\alpha+1}-\phi_{\alpha-1}),
\end{equation}
their relative phases can be gauged away.
On the other hand,
since the sum $\sum_{i=1}^3 \theta_i$ is invariant under these rotation,
the common  phase of $Z_i$
cannot be gauged away generally.
We can only rotate the common phase
of $Z_i$ by $2 \pi /3$, which corresponds to changing the basis of
$3 \times 3$ matrices.
Writing the above solution (\ref{sol1A}) and (\ref{sol1B})
in terms of $U,V$ expansion, we have
\begin{equation}
 A_{\mu}^{(cl)} =X_{\mu} \otimes {\bf 1}_{3 \times 3}, \ \ \
B_{i}^{(cl)} =  Z_i \otimes V.
\label{sol1UV}
\end{equation}
\par
The second type of solution forms a singlet under ${\bf Z}_3$ rotation
in ${\bf C}^3$ plane, and $B_i$ has vanishing vev.
In this case, $A_{\mu}^{(cl)}$ can be an arbitrary diagonal matrix
\begin{equation}
 A_{\mu}^{(cl)}
=\left( \begin{array}{ccc}
  X_{\mu}^{(1)}  & & \\
    & X_{\mu}^{(2)} & \\
    & & X_{\mu}^{(3)}
\end{array}
\right) = \sum_{a=0}^2 X_{\mu}^{a} \otimes U^a.
\label{sol2UV}
\end{equation}
This type of solutions corresponds to the Coulomb branch
in the super Yang-Mills.
These solutions are restricted on the orbifold fixed point and cannot
move into ${\bf C}^3$ direction.
This branch is connected with the Higgs
branch solutions at
$Z_i=0$ and $X_{\mu}^{(1)}=X_{\mu}^{(2)}=X_{\mu}^{(3)}$.
When the Higgs branch solutions come to the orbifold fixed point,
three images with the same four-dimensional coordinates are
liberated, and become possible to move into $d=4$ direction
independently from each other .
\par
We then consider solutions with non-vanishing commutators
$[A_I, A_J] \neq 0$.
The first type (\ref{sol1UV}) with the following commutators
satisfies the equations of motion $[A^I, [A_I,A_J]] =0$:
\begin{equation}
[X_{\mu}, X_{\nu}] = i C_{\mu \nu}{\bf 1}_{N \times N}, \
[Z_i, Z_j^{\dagger}]= D_{ij} {\bf 1}_{N \times N}, \
[Z_i, Z_j] =E_{ij} {\bf 1}_{N \times N},
[X_\mu, Z_i]=F_{\mu,i}{\bf 1}_{N \times N},
\label{Dbrane}
\end{equation}
where $C_{\mu \nu}$, $D_{ij}$, $E_{ij}$ and $F_{\mu,i}$
are c-number coefficients.
If $E_{ij}=F_{\mu,i}=0$, this solution becomes half BPS-saturated and
preserves half of the supersymmetries.
This solution corresponds to an ordinary D-brane.
\par
The second type (\ref{sol2UV}) is a classical solution if
the commutator is proportional to a unit matrix
\begin{equation}
[X_{\mu}^{(\alpha)}, X_{\nu}^{(\alpha)}]
= i C_{\mu \nu}^{(\alpha)} {\bf 1}_{N \times N}.
\end{equation}
This solution is also half BPS-saturated if the coefficient
$C_{\mu \nu}^{(\alpha)}$ is independent of
the block index $(\alpha)$.
An example is
\begin{equation}
X_{0}^{(\alpha)} = \hat{p}, \ X_1^{(\alpha)}= \hat{q},
\ X_2^{(\alpha)}=x_2^{(\alpha)} {\bf 1}_{N \times N},
\ X_3^{(\alpha)}=x_3^{(\alpha)} {\bf 1}_{N \times N}
\label{FDbrane}
\end{equation}
where $[\hat{p}, \hat{q}] =-i$.
It describes three fractional
D-strings whose $x_2$ and $x_3$ coordinates
can be different.  They are constrained in the $z_i=0$ plane.
\subsection{Noncommutative Orbifold Field Theory}
Expanding our orbifold matrix model around the classical
solutions (\ref{Dbrane}), we can obtain
noncommutative field theory in the orbifold background.
Here we set $Z_i=0$, for simplicity.
We write the classical solution $X_{\mu}$
as $\hat{x}_{\mu}$
to emphasise the noncommutativity of coordinates.
Matrices are expanded around this noncommutative background
as
\begin{eqnarray}
 A_{\mu} &=&   \left(
 \begin{array}{ccc}
  \hat{x}_{\mu}+ \tilde{A}_{\mu}^{(1)}  & & \\
    &  \hat{x}_{\mu}+\tilde{A}_{\mu}^{(2)} & \\
    & &  \hat{x}_{\mu}+\tilde{A}_{\mu}^{(3)}
 \end{array}
 \right).
\end{eqnarray}
All other matrices do not have classical background and are
considered as fluctuating fields.
All the fluctuating fields are written in terms of noncommutative
plane waves in  four dimensions
as
\beq
\tilde{A}_{\mu}^a = \sum_k \tilde{A}_{\mu}^a (k)
\exp(i k_{\mu} \hat{x}_{\mu}).
\eeq
By inserting these expansions and adopting the same technique
 to derive noncommutative field theory from IIB matrix model
\cite{NCMM},
the orbifold matrix model  becomes noncommutative orbifold field
theory with ${\cal N}=1$ supersymmetry in $d=4$.
\footnote{
In these cases, $\hat{x}_\mu$
always enters in the action in a form
of commutators.
But if the background is also extended to the ${\bf C}^3$
direction and
$B_i$ is expanded around noncommutative background
$\hat{z}_i$, the same technique cannot be applied
to obtain noncommutative field theory,
because we are faced with a problem that the operator
$\hat{z}_i$ enters into the action
in a different form from commutators.
Typically it is like
$(\hat{z}_i \tilde{A}_{\mu}^{a} -
\omega^a \tilde{A}_{\mu}^{a} \hat{z}_i)$
where the fluctuating fields are expanded in terms of the $U,V$
expansion. Hence, if we expand the fluctuating fields similarly,
we cannot obtain an ordinary kinetic term.
But by deforming the expansion basis of matrices  from the
ordinary noncommutative plane waves,
we may be able to obtain well behaved field theory.
A simple example of deformed plane wave is
$M=\exp(iw \hat{a}^{\dagger}) q^{a^{\dagger}a} \exp(i\bar{w} \hat{a})$.
This satisfies $a M - q Ma=iw M$ where $[a,a^{\dagger}]=1.$
Deformation of plane waves will be related
to the resolution of orbifold singularity by the
introduction of noncommutativity.
Detailed analysis is now under investigation.
\label{singularity}
}
\par
If the background is
replaced by a tensor product
$\hat{x}_{\mu} \otimes {\bf 1}_{n \times n}$,
the resulting noncommutative field theory acquires $U(n)^3$ gauge
symmetry.
This model  contains $d=4$ vector multiplet ($A_{\mu}^a$ and
$\psi_0^a$) in an adjoint representation of $U(n)^3$ and three chiral
multiplets ($B_i^a$ and $\psi_i^a$) in bi-fundamental representations.
It is a chiral four dimensional theory.
We may consider the number of  chiral multiplets, three, as the number of
generations (species) \cite{ishibashi}.
The global symmetry $U(3)$, which is a subgroup of $SO(6)$ part in
$SO(9,1)$, becomes the horizontal symmetry among generations.
 Phenomenological analysis of this model will be
left for future work (see also \cite{pheno}).
\par
We then consider  classical solutions (\ref{FDbrane}).
In this case, if the positions of three fractional D-strings are
largely separated, chiral multiplets $B_i$ and $\psi_i$
acquire large masses
proportional to the distances, and only the vector multiplet
survives in low energy. The low energy fields contain the same
field contents as the $d=4$ super Yang-Mills gauge theory.
Expanding around the  D-string solution (\ref{FDbrane}),
the effective theory becomes  $d=2$ noncommutative
gauge theory obtained from the reduced matrix model of $d=4$
SYM theory. This is consistent
at least qualitatively with the fact that fractional
branes have charges of RR fields in twisted sector, which propagate
in $d=4$ subspace of $d=10$.

\subsection{Effective Action and Interaction between Diagonal Blocks}
We now calculate the one-loop effective action of orbifold matrix model
between two diagonal blocks by the background field method.
To fix gauge invariance in the perturbative calculation,
we add the gauge fixing term
\begin{equation}
 S_{\mbox{gf}} = -\frac{1}{g^2} Tr(\frac{1}{2}
[A_I^{cl}, \tilde{A}^I]^2 +[A_I^{cl},b][A^{I,cl},c]),
\end{equation}
where $b$ and $c$ are ghost fields and $A_I^{cl}$ includes
background configurations of bosonic matrices.
When we expand
\begin{equation}
A_{\mu} = A_{\mu}^{cl} + \tilde{A}_{\mu}, \
B_i = B_i^{cl} + \tilde{B}_i,
\end{equation}
the quadratic part of the quantum fluctuation
in  the bosonic action becomes
\begin{eqnarray}
 \tilde{S}_{b,2} &=& -\frac{1}{2 g^2}
Tr([A_I^{cl},\tilde{A}_J][A^{I,cl},\tilde{A}^J]
+ 2 [A_I^{cl},A_J^{cl}][\tilde{A}^I, \tilde{A}^J])
\nonumber \\
&=&   -\frac{1}{2 g^2}
Tr ( [A_{\mu}^{cl}, \tilde{A}_{\nu}]^2 + [A_{\mu}^{cl}, \tilde{B}_i]
[A^{\mu,cl}, \tilde{B}_i^{\dagger}]
+ [B_i^{cl}, \tilde{A}_{\mu}][B_i^{cl \dagger}, \tilde{A}^{\mu}]
\nonumber \\
&&+ \frac{1}{2}([B_i^{cl},\tilde{B}_j][B_i^{cl \dagger},
\tilde{B}_j^{\dagger}]
+ [B_i^{cl}, \tilde{B}_j^{\dagger}][B_i^{cl \dagger}, \tilde{B}_j])
\nonumber \\
&& + 2 ([A_{\mu}^{cl}, A_{\nu}^{cl}][\tilde{A}^{\mu}, \tilde{A}^{\nu}]
+[A_{\mu}^{cl}, B_i^{cl}][\tilde{A}^{\mu}, \tilde{B}_i^{\dagger}]
+ [A_{\mu}^{cl}, B_i^{cl \dagger}][\tilde{A}^{\mu}, \tilde{B}_i])
\nonumber \\
&& +\frac{1}{2}([B_i^{cl}, B_j^{cl}]
[\tilde{B}_i^{\dagger}, \tilde{B}_j^{\dagger}]
+[B_i^{cl \dagger}, Z_j^{cl \dagger}][\tilde{B}_i, \tilde{B}_j]) \n
&&+ [B_i^{cl}, B_j^{cl \dagger}][\tilde{B}_i^{\dagger}, \tilde{B}_j] ).
\label{action2}
\end{eqnarray}

We first consider the background
of the first-type classical
solution (\ref{sol1UV}) with arbitrary $X_{\mu}$ and $Z_{i}$.
Bosonic matrices are expanded as
\begin{equation}
 A_{\mu} = X_{\mu} \otimes {\bf 1}_{3 \times 3} + \tilde{A}_{\mu}^{a}
        \otimes U^a, \ \ \
B_{i} = Z_{i} \otimes V + \tilde{B}_{i}^{a} \otimes U^a V.
\label{bb1}
\end{equation}
Inserting these expansions into the action (\ref{action2})
and taking the trace over $3 \times 3$ matrices constructed
by $U$ and $V$, we have
\begin{eqnarray}
 \tilde{S}_{b,2} &=& -\frac{3}{2g^2}
Tr (
[X_{\mu}, \tilde{A}_{\nu}^a][X^{\mu}, \tilde{A}^{\nu,-a}]
+ [X_{\mu}, \tilde{B}_i^a][X^{\mu}, \tilde{B}_i^{a \dagger}]
\nonumber \\
&& + (Z_i \tilde{A}_{\mu}^a - \omega^a \tilde{A}_{\mu}^a Z_i)
(\omega^{-a}Z_i^{\dagger}\tilde{A}_{\mu}^{-a} -
\tilde{A}_{\mu}^{-a}Z_i^{\dagger} )
\nonumber \\
&& +\frac{1}{2}(
(Z_i \tilde{B}_j^a - \omega^a \tilde{B}_j^a Z_i)
( \omega^{-a} Z_i^{\dagger}\tilde{B}_j^{a \dagger}-
\tilde{B}_j^{a \dagger} Z_i^{\dagger})
\nonumber \\
&& \ \
+(Z_i \tilde{B}_j^{a \dagger} - \omega^{-a} \tilde{B}_j^{a \dagger} Z_i)
(\omega^{a} Z_i^{\dagger}\tilde{B}_j^{a}-
\tilde{B}_j^{a} Z_i^{\dagger})
)
\nonumber \\
&& + 2(
[X_{\mu},X_{\nu}][\tilde{A^{\mu,a}}, \tilde{A}^{\nu,-a}]
\nonumber \\
&& +
[X_{\mu}, Z_i](\omega^{-a} \tilde{A}^{\mu,a}\tilde{B}_i^{a \dagger}
-  \tilde{B}_i^{a \dagger}\tilde{A}^{\mu,a})
+  [X_{\mu}, Z_i^{\dagger}]
( \tilde{A}^{\mu,a}\tilde{B}_i^{-a}
-  \omega^{-a}\tilde{B}_i^{-a}\tilde{A}^{\mu,a}))
\nonumber \\
&&+ \frac{1}{2}(
[Z_i,Z_j] \omega^a  (\tilde{B}_i^{a,\dagger} \tilde{B}_j^{-a,\dagger}
-\tilde{B}_j^{a,\dagger} \tilde{B}_i^{-a,\dagger})
+[Z_i^\dagger,Z_j^\dagger] \omega^a  (\tilde{B}_i^{a} \tilde{B}_j^{-a}
-\tilde{B}_j^{a} \tilde{B}_i^{-a}))
\nonumber \\
&&+  [Z_i,Z_j^{\dagger}]( \tilde{B}_i^{a,\dagger}
\tilde{B}_j^{a}
- \tilde{B}_j^{a} \tilde{B}_i^{a,\dagger}) ).
\label{Sb2}
\end{eqnarray}
Here $Tr$ is a trace over $N \times N$ after taking the $3 \times 3$ part.
Similarly the quadratic part in quantum fluctuation of the fermionic
matrices becomes
\begin{eqnarray}
 \tilde{S}_{f,2} &=& -\frac{3}{2g^2} Tr
( \sum_{i=0}^3 \overline{\psi_i^a} \Gamma^{\mu} [X_{\mu}, \psi_i^a]
+ 2 \sum_{i=1}^3 \overline{(\psi_i^a)^c}
\bar{\Gamma}^{(i)}
(Z_i^{\dagger} \psi_0^{-a} - \omega^a \psi_0^{-a} Z_i^{\dagger})
\nonumber \\
&& + \sum_{i,j,k=1}^3 |\epsilon_{ijk}|\overline{(\psi_i^a)^c}
\Gamma^{(j)} (\omega^{-a} Z_j \psi_k^{-a} - \omega^{a}\psi_k^{-a}Z_j))
+h.c.
\label{Sf2}
\end{eqnarray}
The last term survives due to the special property for the ${\bf Z}_3$
orbifold matrix model, $V^3={\bf 1}_{3 \times 3}.$
\par
We are interested in the induced interactions between
backgrounds that are separated distantly and hence assume that
$X_{\mu}$ and $Z_i$ are of $2\times 2$ block form
\begin{eqnarray}
X_{\mu}= \left( \begin{array}{cc}
 X_{\mu (1)}&  \\
& X_{\mu (2)}
\end{array}  \right), \ \
Z_i= \left( \begin{array}{cc}
Z_{i (1)}&  \\
& Z_{i (2)}
\end{array}  \right).
\label{bg-block}
\end{eqnarray}
The
difference between the c-number parts of two blocks,
$(\sum_{\mu}(x_{\mu(1)}-x_{\mu(2)})^2+
\sum_i |z_{i(1)}-z_{i(2)}|^2 )^{1/2}$
is assumed to be
much larger than the fundamental scale $g^{1/2}$ of the matrix model.
In order to calculate interactions between these two blocks,
we need to integrate over
the fluctuations $\tilde{A}_{\mu}^a$, $\tilde{B}_i^a$
in the off-diagonal blocks;
\begin{eqnarray}
\tilde{A}_{\mu}^a = \left( \begin{array}{cc}
& \tilde{A}_{\mu (12)}^a  \\
\tilde{A}_{\mu (21)}^a  &
\end{array}  \right), \
\tilde{B}_i^a= \left( \begin{array}{cc}
& \tilde{B}_{i (12)}^a   \\
 \tilde{B}_{i (21)}^a &
\end{array}  \right).
\label{fl-block}
\end{eqnarray}
\par
In the quadratic actions (\ref{Sb2}), (\ref{Sf2}),
$a=0$ sector is always decoupled from $a=1,2$ sector.
This $a=0$ part is the same as the quadratic part of the
 IIB matrix model action with a gauge fixing term
around the background
$ A_{\mu} = X_{\mu}, B_i =Z_i. $
Hence integration over $a=0$ sector gives the same contribution
as the block-block interactions in IIB matrix model
(\ref{bbIIB}), which can be interpreted as exchanges of
massless particles in $d=10$ type IIB supergravity.
The typical form is
\begin{equation}
\frac{1}{((x_{\mu(1)}-x_{\mu(2)})^2 + |z_{i(1)}-z_{i(2)}|^2)^4}
\left( tr(f_{\mu \nu (1)} f_{\nu \lambda(1)}) \
 tr(f_{\mu \sigma (2)} f_{\sigma \lambda(2)}) + \cdots
\right).
\label{typicalint}
\end{equation}
The contributions from the other sectors $a=1,2$ can be
evaluated similarly.
First expand further the actions
(\ref{Sb2}), (\ref{Sf2}) in terms of block fields
(\ref{bg-block}), (\ref{fl-block}).
The bosonic part becomes (we drop the coefficient $-3/2g^2$)
\begin{eqnarray}
&&Tr (
(X_{\mu(1)} \tilde{A}_{\nu(12)}^a -\tilde{A}_{\nu(12)}^a X_{\mu(2)})
(X^{\mu}_{(2)} \tilde{A}_{(21)}^{\nu,-a}-\tilde{A}_{(21)}^{\nu,-a}
X^{\mu}_{(1)})
\nonumber \\
&&+ (X_{\mu(1)} \tilde{B}_{i(12)}^a- \tilde{B}_{i(12)}^a X_{\mu(2)}  )
(X^{\mu}_{(2)} (\tilde{B}_{i(12)}^{a})^{\dagger}-
(\tilde{B}_{i(12)}^{a})^{\dagger}
X^{\mu}_{(1)})
\nonumber \\
&& + (Z_{i(1)} \tilde{A}_{\mu(12)}^a -
\omega^a \tilde{A}_{\mu(12)}^a Z_{i(2)})
(\omega^{-a} Z_{i(2)}^{\dagger}\tilde{A}_{\mu(21)}^{-a} -
\tilde{A}_{\mu(21)}^{-a}Z_{i(1)}^{\dagger} )
\nonumber \\
&& +\frac{1}{2}(
(Z_{i(1)} \tilde{B}_{j(12)}^a - \omega^a \tilde{B}_{j(12)}^a Z_{i(2)})
( \omega^{-a} Z_{i(2)}^{\dagger} (\tilde{B}_{j(12)}^{a})^{\dagger}-
(\tilde{B}_{j(12)}^{a})^{\dagger}Z_{i(1)}^{\dagger} )
\nonumber \\
&&
+(Z_{i(1)} (\tilde{B}_{j(21)}^{a})^{\dagger} -
 \omega^{-a}  (\tilde{B}_{j(21)}^{a})^{\dagger} Z_{i(2)})
(\omega^{a} Z_{i(2)}^{\dagger} \tilde{B}_{j(21)}^{a} -
\tilde{B}_{j(21)}^{a}  Z_{i(1)}^{\dagger})
)
\nonumber \\
&& + 2(
[X_{\mu(1)},X_{\nu(1)}](\tilde{A}^{\mu ,a}_{(12)} \tilde{A}^{\nu,-a}_{(21)}
- \tilde{A}^{\nu,-a}_{(12)}  \tilde{A}^{\mu,a}_{(21)}
)
\nonumber \\
&& +
[X_{\mu(1)}, Z_{i(1)}]
(\omega^{-a} \tilde{A}_{(12)}^{\mu,a} (\tilde{B}_{i(12)}^{a})^{\dagger}
-  (\tilde{B}_{i(21)}^{a})^{\dagger}  \tilde{A}_{(21)}^{\mu,a})
\nonumber \\
&&+  [X_{\mu(1)}, Z_{i(1)}^{\dagger}]
( \tilde{A}^{\mu,a}_{(12)}\tilde{B}_{i(21)}^{-a}
-  \omega^{-a}\tilde{B}_{i(12)}^{-a}\tilde{A}_{(21)}^{\mu,a}))
\nonumber \\
&&+ \frac{1}{2}([Z_{i(1)},Z_{j(1)}]
\omega^a ( (\tilde{B}_{i(21)}^{a})^{\dagger}
(\tilde{B}_{j(12)}^{-a})^{\dagger}
- (\tilde{B}_{j(21)}^{a})^{\dagger}
(\tilde{B}_{i(12)}^{-a})^{\dagger})
\nonumber \\
&&+[Z_{i(1)}^\dagger,Z_{j(1)}^\dagger]
\omega^a ( \tilde{B}_{i(21)}^{a}
\tilde{B}_{j(12)}^{-a}
- \tilde{B}_{j(21)}^{a}
\tilde{B}_{i(12)}^{-a}))
\nonumber \\
&&+  [Z_{i(1)},Z_{j(1)}^{\dagger}]( (\tilde{B}_{i(21)}^{a})^{\dagger}
\tilde{B}_{j(21)}^{a}
- \tilde{B}_{j(12)}^{a} (\tilde{B}_{i(12)}^{a})^{\dagger})
\nonumber \\
&&+   ((1) \leftrightarrow (2), (12) \leftrightarrow (21))).
\end{eqnarray}
The fermionic part becomes
\begin{eqnarray}
&&Tr( \sum_{i=0}^3 \overline{\psi_{i(21)}^a} \Gamma^{\mu}
 (X_{\mu(2)} \psi_{i(21)}^a- \psi_{i(21)}^a X_{\mu(1)})
\nonumber \\
&&+2
 \sum_{i=1}^3 \overline{(\psi_{i(12)}^a)^c}
\overline{\Gamma^{(i)}}
(Z_{i(2)}^{\dagger} \psi_{0(21)}^{-a} -
 \omega^a \psi_{0(21)}^{-a} Z_{i(1)}^{\dagger})
\nonumber \\
&& + \sum_{i,j,k=1}^3 |\epsilon_{ijk}|\overline{(\psi_{i(12)}^a)^c}
\Gamma^{(j)} (\omega^{-a} Z_{j(2)} \psi_{k(21)}^{-a} -
 \omega^{a}\psi_{k(21)}^{-a}Z_{j(1)})
)
\nonumber \\
&&+   ((1) \leftrightarrow (2), (12) \leftrightarrow (21))
+h.c.
\end{eqnarray}
Then we can classify the
fluctuating fields into three classes:
\begin{enumerate}
  \item Fields with $a=0$,
  \item $\tilde{A}_{\mu(12)}^1(=(\tilde{A}_{\mu(21)}^2)^\dagger),
\tilde{B}_{i(12)}^{1}, \tilde{B}_{i(21)}^{2}, \tilde{\psi}_{i(12)}^1,
\tilde{\psi}_{i(21)}^2$,
\item $\tilde{A}_{\mu(12)}^2(=(\tilde{A}_{\mu(21)}^1)^\dagger),
\tilde{B}_{i(12)}^{2}, \tilde{B}_{i(21)}^{1}, \tilde{\psi}_{i(12)}^2,
\tilde{\psi}_{i(21)}^1
$.
\end{enumerate}
The quadratic actions
are decoupled into three parts, each of which contains
only the fields in each class.
Furthermore the quadratic action for each class becomes the same
as the $a=0$ case (class 1) when we absorb the ${\bf Z}_3$ phase
$\omega^a$ into $Z_{i(2)}$
and redefine fields as
\begin{equation}
\omega Z_{i(2)} \rightarrow Z_{i(2)}, \ \
\omega \tilde{B}_{i(12)}^{1} \rightarrow \tilde{B}_{i(12)}^{1}, \ \
 \omega \tilde{\psi}_{i(12)}^1 \rightarrow \tilde{\psi}_{i(12)}^1(i=1,2,3),
\end{equation} for class 2 fields, and
\begin{equation}
\omega^2  Z_{i(2)} \rightarrow Z_{i(2)}, \ \
\omega^2 \tilde{B}_{i(12)}^{2} \rightarrow \tilde{B}_{i(12)}^{2},
 \omega^2 \tilde{\psi}_{i(12)}^2 \rightarrow
\tilde{\psi}_{i(12)}^2(i=1,2,3).
\end{equation}
for class 3 fields.
Therefore, if the background is in the
Higgs branch (\ref{bb1}),
we obtain the same block-block interactions as in IIB
matrix model, except for the replacement
of the typical interaction form
(\ref{typicalint}) by
\begin{eqnarray}
\sum_{a=0}^2
\frac{1}{((x_{\mu(1)}-x_{\mu(2)})^2 + |z_{i(1)}- \omega^a z_{i(2)}|^2)^4}
\left( tr(f_{\mu \nu (1)} f_{\nu \lambda(1)}) \
 tr(f_{\mu \sigma (2)} f_{\sigma \lambda(2)})+ \cdots
\right).
\label{typicalintrep}
\end{eqnarray}
The $a \neq 0$ terms can be interpreted as interactions between
the first block and the
${\bf Z}_3$ transformed  mirror images of the second block.
The tensor structure is the same as the IIB result (\ref{bbIIB})
and each term can be interpreted as an interaction mediated by
massless particles of type IIB supergravity in the untwisted sector
in the orbifold background.
\par
In the remainder of this subsection, we consider an effective action
between Coulomb branch solutions  or fractional branes
in the same twisted sector.
In matrix models, each twisted sector corresponds to each diagonal
block matrices in (\ref{sol2UV}).
In string theory,
fractional branes have charges of massless fields
in the twisted sector, and  the induced interaction
is expected to have $d=4$ behaviour instead of  $d=10$  behaviour.
In the Coulomb branch, since the three fractional branes (\ref{FDbrane})
can move freely into four-dimensional space-time, we can put them
 distantly from each other. Then the chiral multiplets
including $B_i$ and $\psi_i$ become massive (mass is proportional to
the distances) and are decoupled in the low energy.
Only the vector multiplet of $A_{\mu}$ and $\psi_0$ survives.
The low energy theory now becomes three decoupled reduced models of
d=4 super Yang-Mills gauge theory.
The effective interaction between fractional branes in the same
twisted sector is therefore governed by this d=4 reduced model.
This behaviour is qualitatively
consistent with the supergravity point of view.
But in the perturbative calculation, the effective interaction is
inversely proportional to fourth power
of the distances (instead of quadratic)
and inconsistent with the expected power of the supergravity
calculation.
It is not certain if we can overcome this difficulty
nonperturbatively.

\setcounter{equation}{0}
\section{Recovery of IIB Matrix Model}
As we mentioned in the introduction, it is meaningful to make
connection between IIB matrix model and the orbifold matrix model.
If we can dynamically generate the orbifold
matrix model from IIB matrix model,
we could  realize four-dimensional chiral fermion  in
IIB matrix model and
we might also obtain some notion of universality class and
background independence in matrix models.
In this section we report their connection in the opposite direction that
IIB matrix model can be obtained from the orbifold matrix model.
\par
In the Higgs phase (\ref{sol1UV}) with
\beqa
A_\mu^{(cl)}&=&X_{\mu}\otimes {\bf 1}_{3\times 3}=0, \n
B_i^{(cl)}&=&Z_i\otimes V=z_i {\bf 1}_{N \times N}\otimes V,
\label{clsol}
\eeqa
where $z_i$ are complex numbers and $|z_i|^2\gg g$,
$a \neq 0$ sector acquires a large mass and only $a=0$
sector survives.
The resulting model becomes the IIB matrix model,
and ten-dimensional Poincare invariance  and
${\cal N}=2$ supersymmetry of IIB matrix model are recovered.
This result is quite natural because in the large $z_i$ limit,
the orbifold singularity will become invisible,
and fields connecting different mirror sectors will have large mass
proportional to the vev $|z_i|$.
\par
We expand the matrices around the classical solution (\ref{clsol}) as
\beqa
A_{\mu} &=& \tilde{A}_{\mu}^{a} \otimes U^a,  \n
B_{i} &=& z_{i}{\bf 1}_{N\times N} \otimes V + \tilde{B}_{i}^{a} \otimes U^a
V, \n
\psi_0 &=& \psi_{0}^{a} \otimes U^a,  \n
\psi_{i} &=& \psi_{i}^{a} \otimes U^a V,
\eeqa
and
insert them into the orbifold matrix model action
(\ref{bosonaction1}).
We then obtain the quadratic term,
\beqa
\tilde{S}_{b,2}&=&-\frac{3}{4g^2}\sum_{a=0,\pm 1}|\omega^a-1|^2 \n
&&\times Tr[-2|z_i|^2\tilde{A}_{\mu}^a \tilde{A}_{\mu}^{-a}
+\frac{1}{2}(-4|z_i|^2\tilde{B}_j^{a\dagger}\tilde{B}_j^{a}
+2z_i z_j^*\tilde{B}_i^{a\dagger}\tilde{B}_j^{a}
-z_i z_j\tilde{B}_i^{a\dagger}\tilde{B}_j^{-a\dagger}
-z_i^* z_j^*\tilde{B}_i^{a}\tilde{B}_j^{-a}
)]\n
&=&\frac{3}{2g^2}\sum_{a=0,\pm 1}|\omega^a-1|^2 z_1^2
Tr[\tilde{A}_\mu^a \tilde{A}_\mu^{-a}
+\sum_{j=2}^3\tilde{B}_j^{a\dagger}\tilde{B}_j^{a}
+\frac{1}{4}(\tilde{B}_1^{a}+\tilde{B}_1^{-a\dagger})
(\tilde{B}_1^{a\dagger}+\tilde{B}_j^{-a})
].
\eeqa
In the second line, we take $z_1$ real and $z_2=z_3=0$ by using the
$U(3)\subset SO(6)$ invariance of the orbifold matrix model.
Similarly, (\ref{fermionaction1}) gives the quadratic fermionic term
\beq
\tilde{S}_{f,2}=-\frac{3}{g^2}\sum_{a=0,\pm 1}(\omega^{-a}-1)z_1
(\overline{\psi_1^{-a}} \Gamma^{(1)} (\psi_0^a)^c
-\omega^a \overline{(\psi_3^{-a})^c}\Gamma^{(1)}\psi_2^a)
+h.c.
\eeq
All of the $a=0$ modes  are massless,
while
 $a\neq 0$ modes acquire large masses
of the order of $z_1$ except
one mode,
$\tilde{B}_1^{a=1}-(\tilde{B}_1^{a=-1})^\dagger$.
This is a Nambu-Goldstone mode associated with the symmetry breaking
$U(N)^3 \to U(N)$ induced by the vev (\ref{clsol}).
Indeed, the vacuum transforms under $U(N)^3$ as
\beq
[z_1 {\bf 1}_{N\times N}\otimes V,i\theta^a \otimes U^a]
=iz_1(\omega^{-a}-1)\theta^a\otimes U^a V,
\eeq
and $B_1^a=iz_1(\omega^{-a}-1)\theta^a$ satisfies
the relation $(B_1^a)^\dagger =-B_1^{-a}$.
As we will show in the remainder of this section,
this Nambu-Goldstone mode disappears from the spectrum
by the Higgs mechanism.

To discuss general structure of the action, we
 call the classical solution $Z$,
 massless modes in $a=0$ sector $L$,
the above Nambu-Goldstone mode in $a\neq 0$ sector $G$,
and  massive modes in $a\neq 0$ sector $M$, in general.
Then the triple and the quartic parts of the quantum fluctuation
in the action have the following form, in general:
\beqa
\tilde{S}_{b,3}&=&\{ZLLL, ZLGG,ZGGG,ZLLM,ZLGM,ZGGM,O(M^2) \},
 \label{b3ji}\\
\tilde{S}_{b,4}&=&\{LLLL,LLGG,LGGG,GGGG,O(M) \},\label{b4ji}\\
\tilde{S}_{f,3}&=&\{LLL,O(M) \}.\label{f3ji}
\eeqa
The $LLLL$ terms in (\ref{b4ji}) and $LLL$ terms in (\ref{f3ji})
are exactly same as the IIB matrix model action.
The $ZLLL$ terms in (\ref{b3ji}) vanish because $Z$ and $L$ commute.
The $ZLLM$ terms in (\ref{b3ji}) vanish after taking a trace of
$3 \times 3$ blocks
because only $M$ is in
$a\neq 0$ sector while $Z$ and $L$ are in $a=0$ sector.

Therefore, if we take unitary gauge, setting $G=0$,
and integrate out the massive mode $M$,
then we obtain the IIB matrix model action.
If the $ZLLM$ terms were non-vanishing, these terms would induce
new $LLLL$ terms, apart form the IIB matrix model action.

We can also see this Higgs mechanism in other gauges than the unitary gauge.
In the large $z$ limit, the broken symmetry $U(N)^3/U(N)$
becomes the translational symmetry of the Nambu-Goldstone mode $G$.
Therefore, the action, (\ref{b3ji}) and (\ref{b4ji}),
should not depend on $G$, after we integrate out $M$.
Indeed, $ZLGG,ZGGG$ terms in (\ref{b3ji}) vanish
because
\beq
Tr[B_1,B_1^\dagger]^2 \to
6z_1\sum_{a=0,\pm 1}(1-\omega^a)
Tr(\tilde{B}_1^a+\tilde{B}_1^{-a\dagger})
[B_1,B_1^\dagger]^{-a},
\eeq
and this kind of terms must include $M$.
Furthermore, after $M$ is integrated out ,
the $ZLGM,ZGGM$ terms in (\ref{b3ji}) induce
$LLGG,LGGG,GGGG$ terms, which cancel those terms in (\ref{b4ji}).
This cancellation is easy to see in
the $ZZMM,ZGGM,GGGG$ terms:
\beqa
S_b &\to&
\frac{3}{4g^2}|1-\omega|^2
Tr[2z_1^2M_1^\dagger M_1
-\frac{1}{\sqrt{2}}z_1(M_1G^2+M_1^\dagger G^{\dagger2})
+\frac{1}{4}G^{\dagger2}G^2] \n
&=&\frac{3}{4g^2}|1-\omega|^2 2
Tr(z_1 M_1-\frac{1}{2\sqrt{2}}G^{\dagger 2})
(z_1 M_1^\dagger-\frac{1}{2\sqrt{2}}G^{2}),
\eeqa
where
\beq
G=\frac{\tilde{B}_1^1-\tilde{B}_1^{-1\dagger}}{\sqrt{2}},
M_1=\frac{\tilde{B}_1^1+\tilde{B}_1^{-1\dagger}}{\sqrt{2}}.
\eeq

In this way, in the large $z_i$ limit,
the Nambu-Goldstone mode disappears from the spectrum,
only $a=0$ sector survive,
and the IIB matrix model is recovered from
the orbifold matrix model .
\par
Finally in this section, we will shortly discuss the vacuum structure of
the orbifold matrix model.
We have shown that the orbifold matrix model is reduced
to IIB matrix model in the Higgs branch with a large
scalar vev.
Is this vacuum dynamically stable?
In field theory, this vacuum preserves half of supersymmetry
and it is perturbatively stable.
In the case of matrix model, we can discuss more about
the stability of the vacuum.
As studied in \cite{AIKKT}, eigenvalues of bosonic matrices
in IIB matrix model tend to gather and no single eigenvalue
can escape from the others.
The dynamics of the eigenvalues in orbifold matrix model will also
share the same property and then three mirror images of each eigenvalue
will be bounded around the fixed point.
If this is true, the vacuum with large
vev in the Higgs branch is not realized dynamically
and IIB matrix model is not recovered dynamically from the
orbifold matrix model.
Instead the nonperturbative vacuum of the orbifold matrix model
will be such that the orbifold singularity cannot be neglected
and have phenomenologically more interesting properties.
It is more interesting if
we can show that
the  orbifold matrix model can be
dynamically generated from IIB matrix model.
We want to come back to this problem.
\setcounter{equation}{0}
\section{Conclusions and Discussions}
In this paper, we have studied  matrix models in orbifold
background, in particular, ${\bf C}^3/{\bf Z}_3$ orbifold with ${\cal N}=1$
four-dimensional supersymmetry.
We find two types of classical solutions, the Higgs branch solution
and the Coulomb branch solution.
When the solutions have noncommutativity, they correspond
to ordinary D-branes and fractional branes respectively.
We then calculate the one-loop effective action
around these solutions and show that in the former case
we can correctly reproduce the supergravity result
in the orbifold background.
In the latter case, although we can obtain qualitatively
consistent picture with the string theory, perturbative
calculation does not reproduce the supergravity result.
Finally we show that this model includes IIB matrix model
in a special point of the perturbative moduli
in a sense that the orbifold matrix model is reduced to
IIB matrix model when the scalar fields $B_i$ have  large
expectation values in the Higgs branch.
\par
There are several issues that are not studied in this paper.
One of them is whether the orbifold matrix model is
equivalent to the IIB matrix model.
As we discussed in the introduction, various matrix models
might be related to each other.
Since IIB matrix model is obtained from the orbifold matrix model,
we now want to show the reverse.
That is to say,
it is desirable if we can obtain the
orbifold matrix model from IIB matrix model
by some mechanism such as condensations of the fields.
When we simply consider vacuum expectation values
of scalar fields $B_i$ in IIB matrix model of size $3N \times 3N$,
we cannot obtain four dimensional chiral theory.
We have to violate parity and
it is not yet known how we can twist
the background to make the lower dimensional theory
chiral. Perhaps we have to extend the size of IIB matrix model
and consider parity violating condensation outside of the
$3N \times 3N$ matrix.
These are under investigations  together with
other mechanisms  to generate four
dimensional chiral fermions from IIB matrix model dynamically.
\par
Another issue is resolution of the orbifold singularity
by noncommutativity. As we briefly commented in the footnote
\ref{singularity}, when we extend the space-time
into the six dimensional transverse space,
there is a singularity in the extended space.
If we simply apply the same technique for deriving noncommutative
field theories from matrix models as that used in IIB matrix model,
we cannot obtain sensible field theory since the noncommutative
coordinates enter into the action directly without commutators.
But this problem will be overcome by adopting different type
of noncommutative plane waves and this will be related to
the resolution of singularity  due to the noncommutativity
of space. We discuss this problem in our future paper.

\begin{center} \begin{large}
Acknowledgements
\end{large} \end{center}
The work (H. Aoki and S. Iso) is supported
in part by the Grant-in-Aid for Scientific
Research from the Ministry of Education, Science and Culture of Japan.
We thank N.Ishibashi for informing us of his thesis
and J. Nishimura for discussions
on dynamics of orbifold matrix models.

\end{document}